\documentstyle[a4,12pt]{article}
\newcommand{\be}{\begin{equation}}
\newcommand{\nn}{\nonumber}
\newcommand{\bea}{\begin{eqnarray}}
\newcommand{\eea}{\end{eqnarray}}
\newcommand{\ba}{\begin{array}}
\newcommand{\ea}{\end{array}}
\newcommand{\ee}{\end{equation}}

\expandafter\ifx\csname mathbbm\endcsname\relax

\else

\fi \textheight 22cm \textwidth 15cm \topmargin 1mm \oddsidemargin
5mm \evensidemargin 5mm

\newcommand{\beas}{\begin{eqnarray*}}
\newcommand{\eeas}{\end{eqnarray*}}
\newcommand{\bes}{\begin{equation*}}
\newcommand{\ees}{\end{equation*}}
\newcommand{\dir}{\not\!\!{D}}

\newcommand{\lf}{\left}
\newcommand{\ri}{\right}
\newcommand{\f}{\frac}

\def\cL{{\cal L}}

\def\tr           {\mbox{\rm tr}\,}

\def\i2           {\mbox{$\frac{i}{2}$}}

\def\al           {\alpha}
\def\ald           {{\dot {\alpha}}}

\def\bet           {\beta}

\def\del           {\delta}

\def\ep           {\epsilon}

\def\ga           {\gamma}
\def\PB            {{\bar \Phi}}

\def\gad           {{\dot \gamma}}

\def\la           {\lambda}
\def\lab          {\bar \la}

\def\ph           {\phi}

\def\ps           {\psi}
\def\psb          {\bar {\psi}}

\def\fb          {\bar {F}}

\def\rh           {\rho}

\def\si           {\sigma}

\def\sib          {{\bar \sigma}}

\def\th{\theta}

\def\pl           {\partial}

\def\phb           {{\bar{\phi}}}

\def\thb        {{\bar {\theta}}}

\begin{document}

\begin{titlepage}
\hfill \vbox{
    \halign{#\hfil         \cr
           IPM/P-2005/090 \cr
           %SU-ITP-02/10 \cr
           } % end of \halign
      }  % end of \vbox
%\vspace*{20mm}
\begin{center}
{\LARGE {Nonanticommutative Deformation of ${\cal N}=4$ SYM \\
\vspace{1.5mm} 
Theory: The Myers Effect and Vacuum States}}

\vspace*{15mm} \vspace*{1mm} {Reza Abbaspur$^a$  
and Ali Imaanpur$^{a,b}$}

\vspace*{1cm}

{\it $^a$ Institute for Studies in Theoretical Physics and Mathematics (IPM)\\
P.O. Box 19395-5531, Tehran, Iran\\
Email: abbaspur@ipm.ir \\
\vspace*{1mm}
$^b$ Department of Physics, School of Sciences \\
Tarbiat Modares University, P.O. Box 14155-4838, Tehran, Iran\\
Email: aimaanpu@theory.ipm.ac.ir} \\
\vspace*{1cm}

\end{center}

\begin{abstract}
We propose a deformation of ${\cal N}=4$ SYM theory induced by nonanticommutative 
star product. The deformation introduces new bosonic terms which we identify 
with the corresponding Myers terms of a stack of D3-branes in the presence of a 
five-form RR flux. We take this as an indication that the deformed lagrangian describes 
D3-branes in such a background. The vacuum states of the theory are also examined. 
In a specific case where the $U(1)$ part of the gauge field is nonvanishing the (anti)holomorphic transverse coordinates of the brane sit on a fuzzy two-sphere. For a supersymmetric vacuum the antiholomorphic coordinates must necessarily commute. However, 
we also encounter non-supersymmetric vacua for which the antiholomorphic coordinates 
do not commute.    
\end{abstract}

\end{titlepage}

\section{Introduction}

The study of supersymmetric D-branes in the background of a RR flux has revealed new structures on the corresponding superspace. In particular, it turns out that in this background the coordinates of the superspace on the brane do not (anti)commute 
with each other \cite{CVAFA, SEI}.\footnote{ For earlier works on nonanticommutative superspace see \cite{CAS}-\cite{GRASSI}.} Interestingly, this is the sole effect of the background, and hence, as far as the dynamics of D-branes is concerned, one can 
basically ignore the background fields and in effect assume that the coordinates in superspace {\em do not} (anti)commute. The nice thing is that even with such nonanticommuting coordinates one can still construct a super Yang-Mills theory preserving half the ${\cal N}=1$ supersymmetry  \cite{SEI}. If, however, one insists on preserving the whole supersymmetry, then as shown by Ooguri and Vafa \cite{CVAFA}, one further needs to deform the anticommutation relation between the spinor fields on the worldvolume of the brane. The resulting ${\cal N}=1/2$ SYM theory and its generalizations have been extensively studied, see \cite{SEI2}-\cite{C}, for instance.  

In the present work we provide a setting for the study of D3-branes in a graviphoton background. As said above, the graviphoton background introduces a new structure on the superspace coordinates. Accordingly, one needs to refine the superfields definitions, and 
in writing the lagrangian use star products instead of ordinary products. Explicitly, to write an effective lagrangian for D3-branes we proceed  as follows. First, we write the ${\cal N}=4$ lagrangian in terms of ${\cal N}=1$ superfields. The superfields are adapted according to the nonanticommutative nature of the superspace. And finally we use the corresponding nonanticommutative star product in between the superfields. 

There exists, however, a direct way of writing the effective lagrangian and 
checking whether the above construction is consistent. In so doing, we first note that the graviphton flux $C_{\mu\nu}$ is coming from a ten-dimensional five-form RR flux $C_{\mu\nu ijk}$ upon compactification to four dimensions. On the other hand, it is known that how D3-branes respond to this flux; it is through the Chern-Simons action and the Myers terms. For a particular choice of a five-form flux with a zero energy momentum tensor this term has been calculated in \cite{SOHH}. Here, upon a nonanticommutative deformation of ${\cal N}=4$ SYM theory,\footnote{As the ${\cal N}=4$ supercharges carry internal 
$SU(4)$ indices, one can think of some more general deformations of the supersymmetry algebra. For instance, in the case of ${\cal N}=2$ supersymmetry, variant deformations have been considered in \cite{FER}.} we show that the same Myers terms are reproduced. Though, the fermionic terms as well as the supersymmetry transformations will be different than the ones in \cite{SOHH}. Having derived the lagrangian, we examine the vacuum states of the 
theory.  
In the absence of the fermionic fields the vacuum states are the same as those in ordinary 
${\cal N}=4$ theory. In particular, since the theory is defined on Euclidean space,  
we can have new configurations where the {\em holomorphic} scalars, $\ph_i$'s, obey an $SU(2)$ algebra forming a fuzzy two-sphere. Being a vacuum state, the whole configuration will have a zero action. However, in the deformed theory it is also possible to have supersymmetric vacua where a fermionic field ($\lab$) and the gauge field $A_\mu$ are  
nonzero.  Furthermore, we will see that there are vacuum configurations which break supersymmetry. These are characterized by nonzero $U(1)$ connections together with  noncommuting {\em antiholomorphic} constant scalar fields.       

%\cite{CAS, SCH, BOU, KLEMM, GRASSI}.   
The organization of this paper is as follows. In the next section, we begin with the 
preliminaries of the nonanticommutative superspace, and adapt the superfield definitions 
accordingly. We then write the ${\cal N}=4$ lagrangian in terms of the ${\cal N}=1$ superfield language, and use the star product to multiply the superfields. This defines a 
nonanticommutative deformation of ${\cal N}=4$ theory. In section 3, we discuss the 
bosonic terms of the lagrangian which linearly depend on the deformation parameter. 
These terms are precisely the Myers terms appearing in the effective lagrangian of 
multiple D3-branes in the background of a five-form RR flux. In section 4, we discuss the vacuum states of the deformed theory. When the self-dual part of the gauge field strength is nonzero, we will argue how the commutation relation of (anti)holomorphic 
scalars are deformed to that of coordinates of a fuzzy two-sphere. The summary and conclusions are brought in the last section.

\section{Nonanticommutative Deformation of ${\cal N}=4$ SYM}

As mentioned in the Introduction, there are two approaches to study the dynamics of 
D-branes in the graviphton background field. Either one can take care of the background by 
adding appropriate Chern-Simons terms to the DBI action using the Myers prescription 
\cite{Myers, SOHH}. Or alternatively, write the ${\cal N}=4$ theory in terms of ${\cal N}=1$ superfields and use the nonanticommmutative star product. In this section we follow the latter approach, and show that it leads to the correct Myers terms for a stack of 
D3-branes in a five-form flux. In this way, we propose a lagrangian which describes D3-branes in the corresponding graviphoton background. Note that a similar equivalence in the description of D-branes in a Kalb-Ramond $B$ background occurs; one can either introduce the $B$ field directly into the DBI action, or instead, introduce it through 
an appropriate star product between the fields \cite{SW}. 

\subsection{Preliminaries}
To begin with, let us recall the construction of the nonanticommutative superspace in \cite{SEI} where the $\th$ coordinates satisfy the following anticommutation 
relation
\be
\{\th^\al , \th^\bet \}= C^{\al\bet}\, ,
\ee
for $C^{\al\bet}$ a constant symmetric matrix. This, however, requires an ordering for 
the product of functions of $\th$. We choose then to define 
\be
f(\th) * g(\th) =f(\th)\exp \lf( -\f{C^{\al\bet}}{2}\f{\stackrel{\leftarrow}{\pl}}{\pl\th^\al}
\f{\stackrel{\rightarrow}{\pl}}
{\pl\th^\bet} 
\ri)g(\th)\, . \label{RULE1}
\ee
For the chiral multiplet we have
\be
\Phi (y,\th ) = \ph (y) +{\sqrt 2}\th \ps(y) +\th\th F(y)\, .
\ee
However, for the antichiral multiplet and $V$ we choose 
\bea
\PB ({\bar y},\thb ) &=& {\bar \ph} ({\bar y}) +{\sqrt 2}\thb \psb({\bar y})\nn \\
 &+& \thb\thb
\lf({\bar F}({\bar y}) +\f{i}{2}C^{\mu\nu}\{F_{\mu\nu} , \phb \} +iC^{\mu\nu}
\lf\{ A_\nu , D_\mu \phb -\f{i}{4}[A_\mu , \phb ] \ri\}({\bar y})\ri) \nn \\
V(y,\th,\thb)&=& -\th\si^\mu\thb A_\mu(y) + i\th\th\thb\lab(y) -i\thb\thb\th^\al 
\lf(\la_\al (y)+\f{1}{4}\ep_{\al\bet} C^{\bet\ga}\si^\mu_{\ga\gad}\{\lab^\gad 
, A^\mu\}\ri) \nn \\
&+&\f{1}{2}\th\th\thb\thb\lf (D(y) -i\pl_\mu A_\mu(y) \ri)\, ,\label{ANTI}
\eea
where
\be
C^{\mu\nu}\equiv C^{\al\bet}\ep_{\bet\ga}(\si^{\mu\nu})^{\ \ga}_\al\, .
\ee

Also note that $y$ and ${\overline y}$ are related through
\be
{\overline y}^\mu =y^\mu -2i\th\si^\mu\thb\, , 
\ee
together with
\be
[y^\mu , y^\nu] = [y^\mu , \th^\al]=[y^\mu , \thb^\ald]=0\, 
\ee
and thus
\be
[{\overline y}^\mu ,{\overline y}^\nu] = 4\thb\thb C^{\mu\nu}\, .
\ee
Since $\bar{ y}$ coordinates do not commute, for the antichiral superfields 
we define the following star product:
\bea
\PB_1({\bar y},\thb )* \PB_2({\bar y},\thb )&=& \PB_1({\bar y},\thb )
\exp \lf( 2\thb\thb C^{\mu\nu}\f{\stackrel{\leftarrow}{\pl}}{\pl{\bar y}^\mu}
\f{\stackrel{\rightarrow}{\pl}}
{\pl{\bar y}^\nu} \ri)\PB_2({\bar y},\thb )\nn \\
&=& \PB_1({\bar y},\thb )\PB_2({\bar y},\thb ) + 2\thb\thb C^{\mu\nu}
\f{\pl}{\pl{\bar y}^\mu}\PB_1({\bar y},\thb )\f{\pl}{\pl{\bar y}^\nu}
\PB_2({\bar y},\thb )\, . \label{RULE2}
\eea
Notice that the choice of $\PB$ and $V$ in (\ref{ANTI}) ensures that the gauge transformations take the canonical form \cite{SEI, ARAKI}. The chiral and 
antichiral field strength superfields are 
\bea
&& W_\al =-\f{1}{4}\overline {DD} e^{-V}D_\al e^V \, ,\nn \\
&& {\overline W}_\ald=\f{1}{4} DD e^V{\overline D}_\ald e^{-V} \, .
\eea

\subsection{The $C$-deformed Lagrangian}

With these preliminaries on the $C$-deformed superspace, we are now ready to set the stage 
for a particular nonanticommutative version of ${\cal N}=4$ SYM.  A simple prescription for writing the corresponding lagrangian is to express the ${\cal N}=4$ lagrangian in terms of ${\cal N}=1$ superfields and  then use the above star products in between the superfields. In so doing, we recall that the field content of ${\cal N}=4$ theory, in the language of ${\cal N}=1$, consists of a gauge multiplet $W^\al$ together with three chiral multiplets, $\Phi_i$, all in the adjoint representation of the gauge group $U(N)$. So for the deformed lagrangian we get
\bea
\cL &=&  
\int d^2\th\, \tr \lf( W^\al \ast W_\al \ri)+  \int d^2\thb\, \tr \lf({\overline W}_\ald * {\overline W}^\ald\ri)  
  \nn \\
&& + \int d^2\th d^2\thb\, \tr \sum_{i=1}^3 \lf( {\bar \Phi_i} * e^V * \Phi_i * e^{-V} \ri) \nn \\ 
&&+ \f{\sqrt{2}}{2}\int d^2\th\, \tr \lf( \Phi_1 * [\Phi_2 \stackrel{*}{,} \Phi_3] \ri) 
- \f{\sqrt{2}}{2}\int d^2\thb\, \tr \lf( \PB_1 * [\PB_2 \stackrel{\ast}{,} \PB_3] \ri) \, .
\label{L}
\eea 
The above lagrangian is manifestly invariant under the following gauge transformations
\bea
&& e^V \to e^{-i{\overline \Lambda} }* e^V * e^{i \Lambda}\nn \\
&& W^\al \to e^{-i\Lambda }* W^\al * e^{i\Lambda} \nn \\
&& {\overline W}_\ald \to e^{-i{\overline \Lambda} }*{\overline W}_\ald * e^{i{\overline \Lambda}}\nn \\
&& \Phi_i \to e^{-i\Lambda }*\Phi_i * e^{i\Lambda} \nn \\
&& \PB_i \to e^{-i{\overline \Lambda}}*\PB_i * e^{i{\overline \Lambda}}\, .\nn
\eea
Here $\Lambda$ and ${\overline \Lambda}$ are the chiral and antichiral superfields, 
respectively. Also note that the superpotential in (\ref{L}) breaks the original $SO(6)$ R-symmetry to an $SO(3)$ subgroup.

Let us now apply the star product rules (\ref{RULE1}) and (\ref{RULE2}) in (\ref{L}) and do the integrals over the odd coordinates of superspace and write down the lagrangian in terms of the component fields
\bea
\cL = &&\!\!\!\!\!\!\tr\, \lf( -\f{1}{4}F^{\mu\nu}F_{\mu\nu} +i\lab\sib^\mu D_\mu\la +\f{1}{2}D^2 
-D^\mu\phb _i D_\mu\ph_i +i\psb_i\sib^\mu D_\mu\ps_i +\fb_iF_i \ri.\nn \\
 &&-\f{i\sqrt{2}}{2}[\phb_i , \ps_i]\la +\f{i\sqrt{2}}{2}[\ph_i , \psb_i]\lab +
 \f{D}{2}[\ph_i , \phb_i]
-\f{i}{2}C^{\mu\nu}F_{\mu\nu}\lab\lab +\f{1}{8}|C|^2(\lab\lab)^2 \nn \\ 
&&+\f{i}{2}C^{\mu\nu}F_{\mu\nu}\{\phb_i , F_i \} 
 -\f{{\sqrt 2}}{2}C^{\al\bet}
\{D_\mu\phb_i , (\si^\mu\lab)_\al\}\ps_{\bet i} -\f{|C|^2}{16}[\phb_i , \lab]
[\lab , F_i ] \nn \\
&&+ \f{\sqrt{2}}{2}\ep_{ijk}\lf( F^i \ph^j  \ph^k -\ph^i \ps^j \ps^k -\f{1}{12}|C|^2\, 
F^iF^jF^k \ri) \label{LL} \\
&& \lf. -\f{\sqrt{2}}{2}\ep_{ijk}\lf( \fb^i \phb^j  \phb^k -\phb^i \psb^j \psb^k  
+\f{2i}{3}C^{\mu\nu}F_{\mu\nu}\phb^i\phb^j\phb^k   +\f{2}{3}C^{\mu\nu}D_\mu\phb^i D_\nu\phb^j\, \phb^k \ri)\ri)\,  \nn
\eea
where $|C|^2=C^{\mu\nu}C_{\mu\nu}$, and $i,j, \ldots =1,2,3$. Note that terms in the 
last two lines are coming from the {\em deformed superpotential}. The covariant derivatives in the last term appear exactly because of the antichiral superfield definition we used in (\ref{ANTI}).

\section{Myers terms}
In this section we are going to examine the bosonic terms of the superpotential. 
In particular, we will see that the  bosonic terms which are linear in $C$ can 
be identified with the Myers terms. Consider a stack of D3-branes in the presence of a five-form RR flux $C_{\mu\nu ijk}$. We choose an RR flux which has a zero energy-momentum tensor and 
thus it has no back reaction on the metric. The RR flux affects the effective action of 
the D3-branes through the Chern-Simons term, and Myers provides the way one has to calculate this term for multiple branes \cite{Myers}. For our particular choice of 
RR flux, this term has been worked out in \cite{SOHH}. Adapting to our conventions 
in here this term reads  
\be
{S}_{CS}=\f{\al'}{24g^2}\ep^{\mu\nu\rh\si} \int C_{\mu\nu ijk}\, 
\tr\lf(-i\phb^i\phb^j\phb^k F_{\rh\si}+2\phb^iD_\rh\phb^jD_\si\phb^k\ri) 
d^4x \, .\label{CS1}
\ee

In the following we show that if we solve for the auxilary fields in (\ref{LL}) and 
take $C_{\mu\nu}\ep_{ijk}\sim C_{\mu\nu ijk}$, then we reproduce the Myers term in (\ref{CS1}). So let us first solve for the auxilary fields, $D, F_i$ and $\fb_i$, using their equations of motion. This yields
\bea
&&D=-\f{1}{2}[\ph_i , \phb_i] \nn \\
&& F_i = \f{\sqrt{2}}{2}\ep_{ijk}\phb^j\phb^k \nn \\
&& {\bar F}_i = -\f{i}{2}C^{\mu\nu}\{F_{\mu\nu} , \phb_i \} 
+\f{|C|^2}{16}\lf\{[\phb_i , \lab] , \lab \ri\} -\f{\sqrt{2}}{2}
\ep_{ijk}\lf(\ph^j\ph^k -\f{|C|^2}{4}F^jF^k \ri) \label{AUX} 
\eea
Plugging back the auxilary fields (\ref{AUX}) into (\ref{LL}) the lagrangian reads
\bea
\cL = &&\!\!\!\!\!\!\tr \lf( -\f{1}{4}F^{\mu\nu}F_{\mu\nu} +i\lab\sib^\mu D_\mu\la 
-D^\mu\phb _i D_\mu\ph_i +i\psb_i\sib^\mu D_\mu\ps_i  \ri.\nn \\
 &&-\f{1}{8}[\ph_i , \phb_i]^2 +\f{1}{4}[\ph_j , \ph_k][\phb^j , \phb^k] 
+\f{\sqrt{2}}{2}\ep_{ijk}\lf(  \phb^i \psb^j \psb^k  -\ph^i \ps^j \ps^k \ri) \nn \\ 
&& -\f{i\sqrt{2}}{2}[\phb_i , \ps_i]\la +\f{i\sqrt{2}}{2}[\ph_i , \psb_i]\lab \nn \\
&&-\f{i}{2}C^{\mu\nu}F_{\mu\nu}\lab\lab +\f{1}{8}|C|^2(\lab\lab)^2 
-\f{{\sqrt 2}}{2}C^{\al\bet}
\{D_\mu\phb_i , (\si^\mu\lab)_\al\}\ps_{\bet i} 
 \nn \\
&& -\f{\sqrt{2}}{32}|C|^2\ep_{ijk}[\phb^i , \lab][\lab , \phb^j\phb^k ]
-\f{1}{48}|C|^2 \ep_{imn}\ep_{jkl}\, \phb^m\phb^n 
\phb^k\phb^l [\phb^i , \phb^j] \nn \\
&& \lf. -\f{\sqrt{2}}{6}\ep_{ijk}\lf(-iC^{\mu\nu}F_{\mu\nu}\phb^i\phb^j\phb^k   +2C^{\mu\nu}D_\mu\phb^i D_\nu\phb^j\, \phb^k \ri)\ri)\, , \label{SL}
\eea
which is invariant under the so-called ${\cal N}=1/2$ supersymmetry transformations:
\bea
&& \del A_\mu = -i\lab\sib_\mu \xi \nn \\
&& \del \la = -\f{i}{2}\xi [\ph_i , \phb_i] +(F_{\mu\nu}+ \f{i}{2}C_{\mu\nu}\lab\lab )\, \si^{\mu\nu}\xi\, , \ \ \ \  
\del\lab =0 \nn \\
&& \del \phi_i =\sqrt{2} \xi \ps_i\, , \ \ \ \ \ \ \ \ \ \ \  \del \phb_i=0 \nn \\
&& \del \ps_i =\xi\, \ep_{ijk}\, \phb^j\phb^k\, , \ \ \ \ \ \ \  
\del \psb_{\ald i}=-i\sqrt{2}\xi^\al\si_{\al\ald}^\mu D_\mu \phb_i 
\, . \label{ST}
\eea
We observe that upon the identification
\be
C_{\mu\nu}\ep_{ijk}=-\f{\al'}{2{\sqrt 2}}C_{\mu\nu ijk}\, ,
\ee
the bosonic terms linear in $C_{\mu\nu}$ of (\ref{SL}) match exactly to the Myers terms in (\ref{CS1}). The conclusion is that the deformation of ${\cal N}=4$ SYM theory induced by the nonanticommutative star product correctly reproduces the Myers terms. This is a further support for taking (\ref{SL}) as the lagrangian of a stack of 
D3-branes in the five-form flux background. 

It is interesting to compare the supersymmetric lagrangian (\ref{SL}) with the 
one constructed in \cite{SOHH}. In \cite{SOHH}, a term quadratic in $C$ was added by 
hand just for supersymmetric completion. Although the quadratic $C$-terms in these two lagrangians look  different, they are both supersymmetric by themselves.    
Except for this, the two lagrangians have the same bosonic part. The $C$-dependent fermionic parts and the supersymmetry transformations are totally different. 
For the supersymmetry transformations, in \cite{SOHH} a deformation of the kind 
$C_{\mu\nu ijk}\phb^i\phb^j\phb^k$ was introduced in $\del \la$, 
whereas in (\ref{ST}) $\del \la$ is deformed through the term $C_{\mu\nu}\lab\lab$. 
So we conclude that the supersymmetric extension of the system is not unique.  
As for the supersymmetry transformations, the fixed points of the super charges 
are not the same. For the model constructed in \cite{SOHH}, the fixed points do not 
change after the deformation with the RR flux. However, as we discuss in the next section, 
the fixed points of the supersymmetry transformations in (\ref{ST}) will be different.    

To see which lagrangian originates from string theory, one needs to generalize 
and extend the Myers method to calculate the quadratic terms in the RR fields as 
well as the fermionic terms. However, as Seiberg, Ooguri and Vafa \cite{SEI, CVAFA} 
have pointed out, the nonanticommutativity arises if we turn on a graviphoton 
background (which in turn comes from a five-form flux upon compactification to four dimensions). Therefore, we expect a string theory calculations would yield (\ref{SL}) 
as the effective lagrangian of D3-branes in this RR background.     
  
\section{Vacuum States}
In this section we will examine the vacuum states of the model. We first argue that the partition function of the model is independent of the deformation parameter $C$. This happens because the undeformed ${\cal N}=4$ SYM theory has an exact $R$-symmetry. On the 
other hand, all the $C$-dependent terms which appear after the deformation have a positive 
$R$ charge, and hence they will have a zero expectation value in the undeformed theory. So 
we conclude that the partition function is invariant under the deformation. This further 
implies that the vacuum energy remains to be zero. This is similar to what happens in 
the Wess-Zumino model \cite{REY22} and pure ${\cal N}=1/2$ SYM theory \cite{SOH}. Apart from ordinary vacuum states of ${\cal N}=4$ SYM theory, in the following, we will see that the deformed theory admits more vacuum states. First we discuss a set of vacua which are invariant under the supersymmetry transformations. Besides such BPS vacua we also encounter zero energy configurations in which supersymmetry is spontaneously broken.

\subsection{Supersymmetric vacua}
To discuss the BPS states of the model, let us first set the fermoins to zero and look at the bosonic configurations for which the variations of the fermionic fields vanish. 
So requiring $\del \la , \del \psi_i$, and $\del \psb_i$ to be zero we obtain
\bea
&& F_{\mu\nu}^+ =0 \, ,\ \ \ \ \ \ \  D_\mu \phb_i=0\, ,\nn \\
&& [\ph_i , \phb^i] =0 \, , \ \ \ \ \ \ \  [\phb^j , \phb^k]=0 \, ,
\eea
for the BPS configurations. These are the ordinary BPS states of ${\cal N}=4$ 
theory, however, note that here $\ph_i$ and $\phb_i$ are independent and  
the commutator $[\ph_i , \ph_j]$ has not been fixed by the supersymmetry transformations. 
Therefore, one can think of BPS states where $\ph_i$'s (satisfying equations of motion) 
are not commuting. 

An interesting case where vacuum solutions of this kind can appear  
is when $F_{\mu\nu}^+$ is a nonvanishing constant. This is only possible if 
the instanton number vanishes and the fermionic field $\lab$ is turned on. 
To see this, first let us take $\lab \neq 0$, $\del \la =0$, which requires 
\be 
F_{\mu\nu}^+ + \f{i}{2}C_{\mu\nu}\lab\lab =0\, ,\label{DEF}
\ee
together with $\dir\lab =0$ coming from the equation of motion for $\la$. 
Finite action solutions to the above {\em deformed instanton equation} have been 
discussed in \cite{INST}. Here, however, we would like to discuss the zero action 
constant solutions of this equation. In the undeformed ${\cal N}=4$ theory 
($C=0$), for a vacuum state we require both the action and the instanton number to vanish 
implying that $F_{\mu\nu}$ must be zero. But in the deformed theory, the extra $C$-dependent term in the instanton equation allows to have vacua where $F_{\mu\nu}$ 
is nonvanishing. In contrast with the instantons, however, these are not localized 
solutions.   

For simplicity, let us take the $U(1)$ part of the gauge 
field and $\lab$ to be the only nonzero components. A solution to Eq. (\ref{DEF}) 
then is a constant $\lab$ and a constant field strength. As this is a vacuum state we further require its 
instanton number to be zero (for example, one can choose $F_{12}$ to be the only nonzero component of the field strength). For this choice, though, we can take $\phi_i$ to be a constant too, $D_\mu\ph_i=\pl_\mu\ph_i=0$, such that its equation of motion reduces 
to
\be
\lf[[\ph_i , \ph_k], \phb^i\ri] =\f{{\sqrt 2}}{16}|C|^2 \ep_{ijk}\lf( 
\f{7}{3}\phb^i\lab_\ald\lab^\ald \phb^j -\lab_\ald\phb^i\lab^\ald\phb^j + \phb^j\lab_\ald\phb^i\lab^\ald \ri) \, ,\label{PHI}
\ee  
where, in deriving the above equation we have used (\ref{DEF}) and the fact that
\bea
[\phb^j , \phb^k]=0\, .\nn
\eea
Further, since only the $U(1)$ part of the $\lab$ is nonzero and $\phb_i$'s commute 
Eq. (\ref{PHI}) simplifies to
\be
\lf[[\ph_i , \ph_k], \phb^i\ri] = 0\, .\label{EQM}
\ee 
For this to be consistent with the Jacobi identity, we need in addition to require 
that
\be
[\ph_i , \phb_j]=0\, .
\ee
 
One solution to ({\ref{EQM}) is of course $[\ph_i , \ph_j]=0$. However, 
we can also have the following new solution:
\be
[\ph_i , \ph_j] = i\al \, \ep_{ijk} \ph_k \, ,\label{FUZZY}
\ee
preserving the $SO(3)$ symmetry of the action. In general, we expect that the equations 
of motion fix the parameter $\al$. However, here $\al$ remains an arbitrary constant parameter of mass dimension one; it is a moduli parameter in the space of supersymmetric vacua. 

Eq. (\ref{FUZZY}) implies that the holomorphic coordinates $\ph_i$'s satisfy an $SU(2)$ algebra and hence take value on a fuzzy two-sphere. To summarize, turning on a 
graviphton background $C_{\mu\nu}$ gives rise to a new supersymmetric vacuum state 
characterized as follows:
\bea
&&F^{+4}_{\mu\nu}+ \f{i}{2}C_{\mu\nu}\lab^4\lab^4=0\, ,\nn \\
&& [\phb_i , \phb_j]=0 \, , \ \ \ \ \ [\ph_i , \phb_j]=0 \, ,  \nn \\
&& [\ph_i , \ph_j] = i\al\, \ep_{ijk}\, \ph_k \, ,
\eea
where the index $4$ refers to the $U(1)$ part of the gauge group. As a typical 
solution one might take $\phb_i$ to lie in the $U(1)$ subalgebra of $U(N)$, 
and the $\ph_i$ take value in the $SU(2)$ subalgebra of $SU(N)$. An interesting aspect of the above solution is that although it contains a constant nonvanishing field strength and noncommuting scalars it does have a zero action. Also note that 
this state is supersymmetric by construction, and one might expect that it is a direct 
consequence of the condition $S=0$. However, notice that the action is not hermitian and 
therefore supersymmetry is not necessarily followed from the vanishing of the action. 
In the next subsection we will provide such an example where the vacuum state breaks the supersymmetry.    

The above configuration of $\ph_i$'s is reminiscent of a BPS vacuum state in 
${\cal N}=1^*$ model. There the deformation is through the mass 
term, and one interprets the configuration as a collection of $N$ D3-branes sitting on a 
fuzzy two-sphere \cite{POL}. From the supergravity side, the mass deformation is equivalent to turning on a 3-form flux in the bulk. The D3-branes, on the other hand, couple magnetically to this 3-from through the Chern-Simons term, and hence it resembles a 
5-brane wrapped on a two-sphere with $N$ units of RR flux flowing out.   

\subsection{Non-Supersymmetric Vacuum States}
There are yet more constant vacuum solutions which look like a fuzzy sphere and can be constructed if $F_{\mu\nu}^4$ is a nonvanishing constant. Although, these states have a 
zero action, surprisingly they turn out to break the supersymmetry. This is mainly because of the extra $C$-dependent terms in the action and the fact that the theory is defined on Euclidean space where the scalars $\ph_i$ and $\phb_i$ are treated independently. 
To start with, consider the simple case of constant bosonic fields with fermions set to zero. Let us first look at the $\phi_r$ equation of motion, it reads
\bea
&&\lf[[\ph_j , \ph_r] , \phb^j \ri]+i{\sqrt 2}\ep_{ijr}C^{\mu\nu}
F_{\mu\nu}^4\phb^i\phb^j \ =\nn \\ 
&&+\f{|C|^2}{24} \ep_{imn}\ep_{rkj}\lf\{ \phb^m\phb^n\phb^k\phb^j\phb^i 
-\phb^k\phb^j\phb^m\phb^n\phb^i 
+\phb^i\phb^k\phb^j\phb^m\phb^n 
-\phb^i\phb^m\phb^n\phb^k\phb^j
\ri.\nn\\
&&\lf. -[\phb^i ,\phb^j]\phb^m\phb^n\phb^k
 +\phb^k[\phb^i, \phb^j]\phb^m\phb^n
+\phb^m\phb^n[\phb^i , \phb^j]\phb^k
-\phb^k\phb^m\phb^n[\phb^i , \phb^j] \ri\} \, .\label{EQPH}
\eea
Now choose the following ansatz for $\phb_i$ and $\phi_i$
\bea
&&[\phb_i , \phb_j]=i\al\, \ep_{ijk}\phb_k \nn \\ 
&&[ \ph_i , \ph_j ]=0 \, ,\label{ANSATZ}
\eea
where $\al$ is a constant parameter to be fixed by the equation of motion. If we plug 
(\ref{ANSATZ}) into (\ref{EQPH}) we obtain
\be
\al^3 =-4{\sqrt 2}\ \f{C\cdot F}{|C|^2} \, ,
\ee
where $F_{\mu\nu}$ is a constant $U(1)$ field strength, and $C\cdot F\equiv C_{\mu\nu}F^{\mu\nu}$. Notice that $[C]=-1$, so that $\al$ has a mass dimension 1.  

Since $\phb_i$'s are $N\times N$ representations of the $SU(2)$ algebra (\ref{ANSATZ}), 
it follows that
\be
\sum_i \phb_i\phb_i = \f{\al^2}{4}(N^2-1)\, .
\ee 
Using this we can calculate the lagrangian density for this classical configuration,  
where only $\phb_i$ and the $U(1)$ connection are nonzero,   
\be
\cL = -\f{1}{4}\tr F_{\mu\nu}F^{\mu\nu} + \f{1}{6}N(N^2-1)\f{(C\cdot F)^2}{|C|^2}\, . 
\label{LLL}
\ee
We now show that there are $U(1)$ connections for which this lagrangian density 
vanishes and thus the configuration represents a vacuum state. Setting (\ref{LLL}) to 
zero, we get
\be
\f{(C\cdot F)^2}{|F|^2|C|^2}=\f{3}{2(N^2-1)}\, ,\label{RR}
\ee
which has always a solution for $N\geq 2$. Of course, for a vacuum state we must 
further require $F_{\mu\nu}$ to have a zero instanton number. For example, let $F_{12}$ and $F_{13}$ be the only nonvanishing components of $F_{\mu\nu}$. For $C_{\mu\nu}$, take the nonzero components to be $C_{12}=C_{34}$, therefore the lagrangian (\ref{LLL}) becomes
\be
\cL = -\f{N}{2}(F_{12}^2 +F_{13}^2) + \f{1}{6}N(N^2-1) F_{12}^2\, .
\ee
which vanishes for
\be
F_{13} = k\, F_{12}\, ,
\ee
with
\be
k= \pm \sqrt{\f{N^2 -4}{3}} \, .
\ee

In contrast to the ordinary ${\cal N}=4$ SYM theory in which we must restrict to zero 
gauge field strength to discuss the vacua, here we can have vacuum configurations of constant $F_{\mu\nu}$ and $\phb$. The $C$-dependent terms allow a cancellation between 
the contributions of these two fields so that we get a zero action. As $\phb_i$'s are 
not commuting, this vacuum is not supersymmetric, though.

\section{Conclusions}
In this work we studied a deformation of ${\cal N}=4$ SYM theory induced by nonanticommutative star product. We worked out the $C$-dependent terms and showed 
that the bosonic linear terms in $C$ can be identified with the Myers terms of a 
stack of D3-branes in a five-form RR flux. This provided a further support that 
a graviphoton background induces a nonanticommutativity on the worldvolume of the 
brane. So the dynamics can be described either directly by taking into account 
the Myers terms, or the background effects on the dynamics can be captured through 
the nonanticommutative star product. We also discussed classical vacuum states of the 
theory. In addition to the ordinary vacua of ${\cal N}=4$ theory, the theory admits 
vacua where (anti)holomorphic scalars do not commute. This happened mainly because 
the $\ph$ and $\phb$ in Euclidean space are treated as independent fields, and the 
fact that we were interested in preserving only the $Q$ supersymmetry.  
Furthermore, as the action is not hermitian, we can also have vacua which break supersymmetry.   

\hspace{30mm}

%\pagebreak
%\hspace{-6mm}{\large \textbf{Acknowledgement}}


\vspace{1.5mm}

\noindent





\begin{thebibliography}{999}

\bibitem{CVAFA}
H. Ooguri, and C. Vafa, {\em The C-Deformation of Gluino and Non-planar 
Diagrams}, Adv.\ Theor.\ Math.\ Phys.\  {\bf 7}, 53 (2003), [arXiv:hep-th/0302109].

\bibitem{SEI} N. Seiberg, 
{\em Noncommutative Superspace, N=1/2 Supersymmetry, Field Theory and String Theory}, 
JHEP 0306 (2003) 010, [arXiv:hep-th/0305248].

\bibitem{CAS} R.~Casalbuoni, {\em Relativity And Supersymmetries},
Phys.\ Lett.\ B {\bf 62}, 49 (1976),
{\em On The Quantization Of Systems With Anticommutating Variables},
Nuovo Cim.\ A {\bf 33}, 115 (1976), {\em The Classical Mechanics For
Bose-Fermi Systems}, Nuovo Cim.\ A {\bf 33}, 389 (1976).

\bibitem{SCH}
J.H. Schwarz, and P. van Nieuwenhuizen, {\em Speculations Concerning a
Fermionic Structure of Space-time}, Lett. Nuovo Cim. 34 (1982) 21.

\bibitem{BOU}
P.~Bouwknegt, J.~G.~McCarthy and P.~van Nieuwenhuizen, {\em Fusing the
coordinates of quantum superspace}, Phys.\ Lett.\ B {\bf 394}, 82 (1997),
[arXiv:hep-th/9611067].

\bibitem{KLEMM}
D. Klemm, S. Penati, and L. Tamassia, {\em Non(anti)commutative Superspace}, 
Class.Quant.Grav. 20 (2003) 2905, [arXiv:hep-th/0104190]. 
%\cite{Cornalba:2002cu}
\bibitem{Cornalba:2002cu}
  L.~Cornalba, M.~S.~Costa and R.~Schiappa,
  {\em D-brane dynamics in constant Ramond-Ramond potentials and  noncommutative
  geometry}, 
  arXiv:hep-th/0209164.
  %%CITATION = HEP-TH 0209164;%
\bibitem{GRASSI}
J. de Boer, P. Grassi, and P. van Nieuwenhuizen, {\em Non-commutative superspace 
from string theory}, Phys.\ Lett.\ B {\bf 574}, 98 (2003), [arXiv:hep-th/0302078]. 
%

%%%%%%%%%%%%%%%%%%%%%%%%%%%%%%%%%%%%%%%%%%%%%%%%%%%%%%%%%%%%%%%%% 
\bibitem{SEI2} N. Berkovits, and N. Seiberg, 
{\em Superstrings in Graviphoton Background and N=1/2+3/2 Supersymmetry}, 
JHEP 0307 (2003) 010, [arXiv:hep-th/0306226].

\bibitem{REZA}R. Abbaspur, {\em Noncommutative Supersymmetry in Two-dimensions}, 
Int. J. Mod. Phys. A18:855-878,2003, [arXiv:hep-th/0110005], {\em Generalized Noncommutative Supersymmetry from a New Gauge Symmetry}, arXiv:hep-th/0206170,   
{\em Scalar Solitons in Non(anti)commutative Superspace}, 
{arXiv:hep-th/0308050}. 

%\cite{Terashima:2003ri}
\bibitem{B}
  S.~Terashima and J.~T.~Yee,
  {\em Comments on noncommutative superspace}, 
  JHEP {\bf 0312}, 053 (2003)
  [arXiv:hep-th/0306237].
  %%CITATION = HEP-TH 0306237;%%
\bibitem{GPR}
  M.~T.~Grisaru, S.~Penati and A.~Romagnoni,
  {\em Two-loop renormalization for nonanticommutative N = 1/2 supersymmetric  WZ
  model}, 
  JHEP {\bf 0308}, 003 (2003)
  [arXiv:hep-th/0307099], A.~Romagnoni,
  %``Renormalizability of N = 1/2 Wess-Zumino model in superspace,''
  JHEP {\bf 0310}, 016 (2003)
  [arXiv:hep-th/0307209], S.~Penati and A.~Romagnoni,
  %``Covariant quantization of N = 1/2 SYM theories and supergauge invariance,''
  JHEP {\bf 0502}, 064 (2005)
  [arXiv:hep-th/0412041].
  %%CITATION = HEP-TH 0412041;%%
\bibitem{B3}
  M.~Alishahiha, A.~Ghodsi and N.~Sadooghi, {\em One-loop perturbative corrections to non(anti)commutativity parameter  of N = 1/2 supersymmetric U(N) gauge theory},
  Nucl.\ Phys.\ B {\bf 691}, 111 (2004)
  [arXiv:hep-th/0309037].
  %%CITATION = HEP-TH 0309037;%%
  \bibitem{A1}
  A.~Sako and T.~Suzuki, {\em Ring structure of SUSY * product and 1/2 SUSY Wess-Zumino model}, 
  Phys.\ Lett.\ B {\bf 582}, 127 (2004)
  [arXiv:hep-th/0309076].
  %%CITATION = HEP-TH 0408226;%%
  %%CITATION = HEP-TH 0309076;%%
  \bibitem{A2}
  B.~Chandrasekhar and A.~Kumar,
  {\em D = 2, N = 2, supersymmetric theories on non(anti)commutative superspace}, 
  JHEP {\bf 0403}, 013 (2004)
  [arXiv:hep-th/0310137].
  %%CITATION = HEP-TH 0408184;%%
  %%CITATION = HEP-TH 0310137;%%
  \bibitem{A3}
  S.~Iso and H.~Umetsu, {\em Gauge theory on noncommutative supersphere from supermatrix model},
  Phys.\ Rev.\ D {\bf 69}, 105003 (2004)
  [arXiv:hep-th/0311005].
  %%CITATION = HEP-TH 0311005;%%
\bibitem{ABC}  A.~Imaanpur and S.~Parvizi,
  {\em N = 1/2 super Yang-Mills theory on Euclidean $AdS_2 \times S^2$}, 
  JHEP {\bf 0407}, 010 (2004)
  [arXiv:hep-th/0403174].
  %%CITATION = HEP-TH 0403174;%%
%%CITATION = HEP-TH 0308171;%%
%\cite{Grisaru:2003fd}
  %\cite{Ihl:2005zd}
\bibitem{Ihl:2005zd} M.~Ihl and C.~Saemann,
  {\em Drinfeld-twisted supersymmetry and non-anticommutative superspace}, 
  arXiv:hep-th/0506057.
  %%CITATION = HEP-TH 0506057;%%
  %\cite{Ivanov:2003te}
\bibitem{Ivanov:2003te}
  E.~Ivanov, O.~Lechtenfeld and B.~Zupnik,
  {\em Nilpotent deformations of N = 2 superspace}, 
  JHEP {\bf 0402}, 012 (2004)
  [arXiv:hep-th/0308012], Nucl.\ Phys.\ B {\bf 707}, 69 (2005)
  [arXiv:hep-th/0408146], 
  S.~Ferrara, E.~Ivanov, O.~Lechtenfeld, E.~Sokatchev and B.~Zupnik,
  %``Non-anticommutative chiral singlet deformation of N = (1,1) gauge
  %theory,''
  Nucl.\ Phys.\ B {\bf 704}, 154 (2005)
  [arXiv:hep-th/0405049].
  %%CITATION = HEP-TH 0405049;%%
  \bibitem{A6}
  T.~Araki, K.~Ito and A.~Ohtsuka, {\em N = 2 supersymmetric U(1) gauge theory in noncommutative harmonic superspace}, 
  JHEP {\bf 0401}, 046 (2004)
  [arXiv:hep-th/0401012], JHEP {\bf 0505}, 074 (2005)
  [arXiv:hep-th/0503224]. 
  %%CITATION = HEP-TH 0503224;%%
  %%CITATION = HEP-TH 0410203;%%
  %%CITATION = HEP-TH 0404250;%%
  %%CITATION = HEP-TH 0401012;%%
  \bibitem{A7}
  C.~Saemann and M.~Wolf,
  {\em Constraint and super Yang-Mills equations on the deformed superspace
  R(h)(4$|$16)}, 
  JHEP {\bf 0403}, 048 (2004)
  [arXiv:hep-th/0401147].
  %%CITATION = HEP-TH 0401147;%%
  \bibitem{A8}
   M.~Billo, M.~Frau, F.~Lonegro and A.~Lerda,
  {\em N = 1/2 quiver gauge theories from open strings with R-R fluxes}, 
  JHEP {\bf 0505}, 047 (2005)
  [arXiv:hep-th/0502084].
  %%CITATION = HEP-TH 0502084;%%
  %%CITATION = HEP-TH 0402160;%%
  \bibitem{A11}
  D.~Mikulovic,
  {\em Seiberg-Witten map for superfields on N = (1/2,0) and N = (1/2,1/2)
  deformed superspace}, 
  JHEP {\bf 0405}, 077 (2004)
  [arXiv:hep-th/0403290].
  %%CITATION = HEP-TH 0403290;%%
  \bibitem{A22}
  S.~V.~Ketov and S.~Sasaki,
  {\em BPS-type equations in the non-anticommutative N = 2 supersymmetric U(1)
  gauge theory}, 
  Phys.\ Lett.\ B {\bf 595}, 530 (2004)
  [arXiv:hep-th/0404119]. 
  %%CITATION = HEP-TH 0407211;%%
  %%CITATION = HEP-TH 0405278;%%
  %%CITATION = HEP-TH 0404119;%%
  
  \bibitem{A33}
  A.~T.~Banin, I.~L.~Buchbinder and N.~G.~Pletnev,
  {\em Chiral effective potential in N = 1/2 non-commutative Wess-Zumino  model},
  JHEP {\bf 0407}, 011 (2004)
  [arXiv:hep-th/0405063], O.~D.~Azorkina, A.~T.~Banin, I.~L.~Buchbinder and N.~G.~Pletnev,
  %``Generic chiral superfield model on nonanticommutative N = 1/2 superspace,''
  Mod.\ Phys.\ Lett.\ A {\bf 20}, 1423 (2005)
  [arXiv:hep-th/0502008], arXiv:hep-th/0509193.
  %%CITATION = HEP-TH 0509193;%%
 \bibitem{A66}
  A.~Gorsky and M.~Shifman,
  {\em Spectral degeneracy in supersymmetric gluodynamics and one-flavor QCD
  related to N = 1/2 SUSY}, 
  Phys.\ Rev.\ D {\bf 71}, 025009 (2005)
  [arXiv:hep-th/0410099].
  %%CITATION = HEP-TH 0410099;%%
  \bibitem{A77}
  L.~G.~Aldrovandi, D.~H.~Correa, F.~A.~Schaposnik and G.~A.~Silva,
  {\em BPS analysis of gauge field - Higgs models in non-anticommutative
  superspace}, 
  Phys.\ Rev.\ D {\bf 71}, 025015 (2005)
  [arXiv:hep-th/0410256].
  %%CITATION = HEP-TH 0410256;%%
  \bibitem{A88}
  T.~Hatanaka, S.~V.~Ketov, Y.~Kobayashi and S.~Sasaki,
  {\em Non-anti-commutative deformation of effective potentials in supersymmetric
  gauge theories}, 
  Nucl.\ Phys.\ B {\bf 716}, 88 (2005)
  [arXiv:hep-th/0502026].
  %%CITATION = HEP-TH 0502026;%%
  \bibitem{A12}
  L.~Alvarez-Gaume and M.~A.~Vazquez-Mozo,
  {\em On nonanticommutative N = 2 sigma-models in two dimensions},
  JHEP {\bf 0504}, 007 (2005)
  [arXiv:hep-th/0503016].
  %%CITATION = HEP-TH 0503016;%%
  \bibitem{A13}
  T.~A.~Ryttov and F.~Sannino,
  {\em Chiral models in noncommutative N = 1/2 four dimensional superspace}, 
  Phys.\ Rev.\ D {\bf 71}, 125004 (2005)
  [arXiv:hep-th/0504104].
  %%CITATION = HEP-TH 0504104;%%
  \bibitem{A14}
   Y.~Kobayashi and S.~Sasaki,
  {\em Non-local Wess-Zumino model on nilpotent noncommutative superspace}, 
  Phys.\ Rev.\ D {\bf 72}, 065015 (2005)
  [arXiv:hep-th/0505011].
  %%CITATION = HEP-TH 0505011;%%
  \bibitem{A15}
  S.~Giombi, R.~Ricci, D.~Robles-Llana and D.~Trancanelli,
  {\em Instantons and matter in N = 1/2 supersymmetric gauge theory}, 
  arXiv:hep-th/0505077.
  %%CITATION = HEP-TH 0505077;%%
  \bibitem{A16}
  C.~S.~Chu and T.~Inami,
  {\em Konishi anomaly and central extension in N = 1/2 supersymmetry}, 
  Nucl.\ Phys.\ B {\bf 725}, 327 (2005)
  [arXiv:hep-th/0505141].
  %%CITATION = HEP-TH 0505141;%%
  \bibitem{A17}
  J.~S.~Cook,
  {\em Gauged Wess-Zumino model in noncommutative Minkowski superspace}, 
  arXiv:hep-th/0505247.
  %%CITATION = HEP-TH 0505247;%%
 \bibitem{A18}
  T.~Araki, T.~Takashima and S.~Watamura,
  {\em On a superfield extension of the ADHM construction and N = 1 super
  instantons}, 
  JHEP {\bf 0508}, 065 (2005)
  [arXiv:hep-th/0506112].
  %%CITATION = HEP-TH 0506112;%%
  \bibitem{C}
  K.~Ito and H.~Nakajima,
  {\em Non(anti)commutative N = 2 supersymmetric U(N) gauge theory and deformed
  instanton equations}, 
  arXiv:hep-th/0508052.
  %%CITATION = HEP-TH 0508052;%%

%%%%%%%%%%%%%%%%%%%%%%%%%%%%%%%%%%%%%%%%%%%%%%%%%%%%%%%%%%%%%%%%%
%\cite{Ferrara:2003xk}
\bibitem{FER} S.~Ferrara and E.~Sokatchev, {\em Non-anticommutative N = 2 
super Yang-Mills theory with singlet deformation}, Phys.\ Lett.\ B {\bf 579}, 
226 (2004), [arXiv:hep-th/0308021], T.~Araki and K.~Ito,
  {\em Singlet deformation and non(anti)commutative N = 2 supersymmetric U(1)
  gauge theory}, Phys.\ Lett.\ B {\bf 595}, 513 (2004)
  [arXiv:hep-th/0404250], T.~Araki, K.~Ito and A.~Ohtsuka,
  {\em Deformed supersymmetry in non(anti)commutative N = 2 supersymmetric U(1)
  gauge theory}, Phys.\ Lett.\ B {\bf 606}, 202 (2005)
  [arXiv:hep-th/0410203].
  %%CITATION = HEP-TH 0410203;%%

%\cite{Imaanpur:2005pd}
\bibitem{SOHH}
A.~Imaanpur, {\em Supersymmetric D3-branes in Five-Form Flux}, 
JHEP {\bf 0503}, 030 (2005), [arXiv:hep-th/0501167].
%%CITATION = HEP-TH 0501167;%%

%\cite{Myers:1999ps}
\bibitem{Myers} R. Myers, {\em Dielectric-branes}, 
JHEP {\bf 9912} (1999) 022, [arXiv:hep-th/9910053].
%%CITATION = HEP-TH 9910053;%%

%\cite{Seiberg:1999vs}
\bibitem{SW} N.~Seiberg and E.~Witten, {\em String theory and noncommutative geometry}, JHEP {\bf 9909}, 032 (1999), [arXiv:hep-th/9908142].
%%CITATION = HEP-TH 9908142;%%

\bibitem{ARAKI}
T. Araki, K. Ito, and A. Ohtsuka, {\em Supersymmetric Gauge Theories on 
Noncommutative Superspace}, Phys.\ Lett.\ B {\bf 573}, 209 (2003), 
[arXiv:hep-th/0307076].

%\cite{Britto:2003aj}
\bibitem{REY22}R.~Britto, B.~Feng and S.~J.~Rey, {\em Deformed superspace, N = 1/2 supersymmetry and (non)renormalization theorems}, JHEP {\bf 0307}, 067 (2003), 
[arXiv:hep-th/0306215].
%%CITATION = HEP-TH 0306215;%%

%\cite{Imaanpur:2003ig}
\bibitem{SOH}
A.~Imaanpur, {\em Comments on gluino condensates in N = 1/2 SYM theory}, 
JHEP {\bf 0312}, 009 (2003), [arXiv:hep-th/0311137].
%%CITATION = HEP-TH 0311137;%%

\bibitem{INST} A.~Imaanpur, {\em On instantons and zero modes of N = 1/2 SYM theory}, 
JHEP {\bf 0309}, 077 (2003), [arXiv:hep-th/0308171], P.~A.~Grassi, R.~Ricci and D.~Robles-Llana, {\em Instanton calculations for N = 1/2 super Yang-Mills theory}, JHEP {\bf 0407}, 065 (2004) [arXiv:hep-th/0311155], 
  R.~Britto, B.~Feng, O.~Lunin and S.~J.~Rey,
  {\em U(N) instantons on N = 1/2 superspace: Exact solution and geometry of
  moduli space}, 
  Phys.\ Rev.\ D {\bf 69}, 126004 (2004)
  [arXiv:hep-th/0311275], 
  M.~Billo, M.~Frau, I.~Pesando and A.~Lerda,
  {\em N = 1/2 gauge theory and its instanton moduli space from open strings  in
  R-R background}, JHEP {\bf 0405}, 023 (2004)
  [arXiv:hep-th/0402160].

%\cite{Polchinski:2000uf}
\bibitem{POL} J.~Polchinski and M.~J.~Strassler, {\em The string dual of a confining four-dimensional gauge theory}, arXiv:hep-th/0003136.
%%CITATION = HEP-TH 0003136;%%







\end{thebibliography}
\end{document}